\documentclass[twocolumn,showpacs,preprintnumbers,amsmath,amssymb,superscriptaddress]{revtex4}
\usepackage[dvips]{graphicx}
\usepackage{graphicx,color}
\usepackage{latexsym,epsfig,bm,times,psfrag,subfigure}
\usepackage{color}
\usepackage{dcolumn}                 
\usepackage[bookmarksnumbered,bookmarksopen,colorlinks,citecolor=blue,linkcolor=blue]{hyperref} %
\usepackage{color, soul}
\setulcolor{red} 
\setstcolor{red} 
\sethlcolor{green} 

\begin{document}
\title{Initial Conditions for Modified DGLAP Evolution of the Modified Fragmentation Functions in Nuclear Medium}

\author{Ning-Bo Chang}
\affiliation{School of Physics, Shandong University, Jinan, Shandong 250100, China}
\affiliation{Key Laboratory of Quark and Lepton Physics (MOE) and Institute of Particle Physics, Central China Normal University, Wuhan 430079, China}
\author{Wei-Tian Deng}
\affiliation{Theory Center, IPNS, KEK, 1-1 Oho, Tsukuba, Ibaraki 305-0801, Japan}

\author{Xin-Nian Wang}
\affiliation{Key Laboratory of Quark and Lepton Physics (MOE) and Institute of Particle Physics, Central China Normal University, Wuhan 430079, China}
\affiliation{Nuclear Science Division Mailstop 70R0319,  Lawrence Berkeley National Laboratory, Berkeley, California 94740, USA}

\date{\today}
\pacs{24.85.+p, 12.38.Bx, 13.87.Ce, 13.60.-r}

\begin{abstract}
Initial conditions are required to solve medium modified DGLAP (mDGLAP) evolution equations for modified fragmentation functions due to multiple scatterings and parton energy loss. Such initial conditions should in principle include energy loss for partons at scale $Q_0$ above which mDGLAP evolution equations can be applied. Several models for the initial condition motivated by induced gluon bremsstrahlung in perturbative QCD are used to calculate the modified fragmentation functions in nuclear medium and to extract the jet transport parameter $\hat q$ from fits to experimental data in deeply inelastic scattering (DIS) off nuclei. The model with a Poisson convolution of multiple gluon emissions is found to provide the overall best $\chi^2$/d.o.f. fit to the HERMES data and gives a value of $\hat q_0 \approx 0.020 \pm 0.005$ GeV$^2$/fm at the center of a large nucleus.
\end{abstract}

\maketitle

\section{Introduction}\label{sec:intro}

When a hard parton passes through a medium, either cold nuclear matter or quark-gluon plasma (QGP), it will lose energy due to multiple scatterings and induced gluon bremsstrahlung. Its fragmentation function (FF) into final hadrons will be modified as compared to that in vacuum.  One can then measure such medium modification of the fragmentation functions or the final hadron spectra to extract medium properties such as the jet transport parameter. The modification in general involves suppression of leading hadrons in deeply inelastic scattering (DIS) off nuclei or high transverse momentum hadron spectra in high-energy heavy-ion collisions. Such phenomena referred to as jet quenching have been the focus of many theoretical \cite{Bjorken:1982tu,Gyulassy:1990ye,Wang:1991xy,Gyulassy:1993hr,Baier:1994bd,Baier:1996sk,Zakharov:1996fv,Gyulassy:2000fs,Gyulassy:2000er,Wiedemann:2000za,Wiedemann:2000tf,Guo:2000nz,Wang:2001ifa,Arnold:2001ba,Arnold:2002ja,Wang:1998ww,Gyulassy:2000gk} and 
experimental studies \cite{Adams:2005dq, Adcox:2004mh,Jacobs:2004qv, Majumder:2010qh,Muller:2012zq} in the last two decades. They have provided important information about the properties of dense medium that is created in high-energy heavy-ion collisions. 

In the latest survey study by the JET Collaboration \cite{Burke:2013yra}, fits to experimental data on the suppression factors of single hadron spectra in high-energy heavy-ion collisions at both the Relativistic Heavy-ion Collider (RHIC) and the Large Hadron Collider (LHC)  indicate values of $\hat q \approx 1.2 \pm 0.3$ and $1.9 \pm 0.7$ GeV$^2$/fm at the center of the most central Au+Au collisions at $\sqrt{s}=200$ GeV/n and Pb+Pb collisions at $\sqrt{s}=2.76$  TeV/n, respectively, at an initial time $\tau_0=0.6$ fm/$c$ for a quark jet with initial energy of 10 GeV. Uncertainties in the extracted values of $\hat q$, though much reduced from previous studies \cite{Bass:2008rv}, are still large and arise mainly from both errors in experimental data on jet quenching measurements and different model implementations of parton energy loss. One of the model implementations is based on the high-twist approach to multiple scatterings and induced gluon radiation \cite{Guo:2000nz,Wang:2001ifa} in which one can 
calculate medium modifications of the fragmentation functions that lead to the observed suppression of final hadron spectra in high-energy heavy-ion collisions. Inclusion of multiple gluon emissions can be achieved through a set of modified Dokshitzer-Gribov-Lipatov-Altarelli-Parisi (DGLAP) evolution equations \cite{Wang:2009qb,Deng:2010xv}. One needs, however, initial conditions for the fragmentation functions at the lowest scale $Q_0$ to calculate modified fragmentation functions at the scale $Q$ of the jet production by solving the modified DGLAP (mDGLAP) evolution equations. Different choices of the initial conditions, which in principle are not calculable in pQCD, contribute to the theoretical uncertainties in the extracted values of the jet transport parameter from experimental data on jet quenching. In this paper, we will use several models for the initial conditions for medium modified fragmentation functions of a quark jet propagating through cold nuclear matter in DIS off large nuclei. We assess the quality of the $\chi^2$/d.o.f. of fits using the calculated  final spectra of leading hadrons with different initial conditions to the experimental data and extract the best values of the jet transport parameter in the cold nuclei.

\section{Modified DGLAP equations and initial conditions for fragmentation functions}\label{sec:ini}

Within the high-twist approach, one can calculate medium modification to the parton fragmentation functions in DIS through higher-twist corrections to the semi-inclusive cross section. The higher-twist corrections can be expressed in terms of medium modified parton fragmentation function  \cite{Guo:2000nz,Wang:2001ifa,Zhang:2003yn,Schafer:2007xh}.  Such an approach to parton energy loss through medium modified fragmentation functions (mFF's)  has been employed to describe suppression of leading hadrons in DIS, nuclear modification of Drell-Yan spectra in p+A collisions \cite{Xing:2011fb} as well as jet quenching in high-energy heavy-ion collisions \cite{Wang:2002ri,Majumder:2004pt,Majumder:2007ae,Chen:2010te,Chen:2011vt}.  In this description of parton energy loss with single gluon emission, medium corrections to the fragmentation functions can become large enough ( long propagation length in very dense medium) to make the modified fragmentation functions (vacuum + medium correction) negative at large fractional momentum $z$. To maintain positivity of mFF's, 
one simply sets them to be zero whenever their values become negative in the large $z$ region.  One solution to the problem of negative mFF's is to include multiple gluon emissions through resummation. This will lead to a medium modified DGLAP (mDGLAP) equations for the mFF's \cite{Wang:2009qb,Deng:2010xv,Majumder:2011uk,Majumder:2013re},
\begin{eqnarray}
 \label{eq: modified DGLAP1}
 \frac{\partial \tilde{D}_q^h(z_h,Q^2)}{\partial \ln Q^2}\hspace{-4pt} &=&\hspace{-4pt}\frac{\alpha_s(Q^2)}{2\pi}\hspace{-4pt} \int_{z_h}^1
 \frac{dz}{z}\left [ \tilde{\gamma}_{q\rightarrow qg}(z,Q^2)\tilde{D}_q^h(\frac{z_h}{z},Q^2)\right. \nonumber \\
 &+&\left.\tilde{\gamma}_{q\rightarrow gq}(z,Q^2)\tilde{D}_g^h(\frac{z_h}{z},Q^2)\right ] ,\\
\label{eq: modified DGLAP2}
  \frac{\partial \tilde{D}_g^h(z_h,Q^2)}{\partial\ln Q^2}\hspace{-4pt}&=&\hspace{-4pt}\frac{\alpha_s(Q^2)}{2\pi} \hspace{-4pt} \int_{z_h}^1
 \frac{dz}{z}\left [ \tilde{\gamma}_{g\rightarrow gg}(z,Q^2)\tilde{D}_g^h(\frac{z_h}{z},Q^2) \right.  \nonumber \\
 &+&\left. \sum_{q=1}^{2n_f}\tilde{\gamma}_{g\rightarrow q\bar q}(z,Q^2)\tilde{D}_q^h(\frac{z_h}{z},Q^2)\right ] ,
\end{eqnarray}
where the modified splitting functions $\tilde \gamma _{a\rightarrow bc}$ are given by the sum of the vacuum ones 
and the medium modification,
\begin{equation}
 \tilde \gamma _{a\rightarrow bc}(z,Q^2)=\gamma_{a\rightarrow bc}(z)+ \Delta \gamma_{a\rightarrow bc}(z,Q^2),
\end{equation}
which can be found in \cite{Schafer:2007xh}.

To solve the mDGLAP equations, one has to provide initial conditions of mFF's at a given scale $Q_0$ which in principle are not calculable in pQCD. The simplest assumption is that these initial conditions take the form of fragmentation functions in the vacuum \cite{Majumder:2011uk,Majumder:2013re}. This {\it vacuum } initial condition assumes that there is no medium interaction and parton energy loss for partons below scale $Q_0$. It understandably underestimates the total parton energy loss and requires larger values of $\hat q$ to fit the experimental data on jet quenching. It also gives stronger $Q^2$ dependence of the medium modification than the experimental data.  In our previous work \cite{Wang:2009qb,Deng:2010xv}, we assumed a model for the initial condition, which is obtained by evolving the vacuum fragmentation functions at scale $Q_0$ according to a set of mDGLAP equations with only medium induced splitting functions from $Q=0$ to $Q_0$. This model for initial 
conditions, which we will refer to as {\it evolved} initial conditions,  tends to overestimate the parton energy loss and our preliminary study shows that it will lead to a wrong momentum dependence of jet quenching in heavy-ion collisions at the LHC \cite{cdw14}. 

In this paper, we will consider another model of multiple gluon emissions below scale $Q_0$, which we will refer to as {\it convoluted} initial conditions.  We introduce a quenching weight $P(\epsilon)$ to represent the probability of a parton losing a fraction $\epsilon=\Delta E/E$ of its energy. The quenching weight is assumed to be given by a Poisson convolution of multiple gluon emissions, each of which is determined by the induced gluon spectrum from a single emission. Such a model has been used in other approaches to parton energy loss in dense medium for both an on-shell or highly virtual parton \cite{Wang:1996yh,Baier:2001yt,Salgado:2002cd,Gyulassy:2001nm}. 

With the assumption that the number of independent induced gluon emissions satisfies the Poisson distribution, 
the probability of fractional energy loss $\epsilon=\Delta E/E$ by a propagating parton with virtuality $Q_0^2$ can 
then be expressed as \cite{Chen:2010te,Wang:2009qb}
\begin{eqnarray}
  P_a(\epsilon,Q_0^2)&=&\sum_{n=0}^\infty \frac{1}{n!}
  \left[ \prod_{i=1}^n \int_0^{1} dz_i \frac{dN_g^a}{dz_i}(Q_0^2)
    \right]
    \delta\left(\epsilon-\sum_{i=1}^n z_i\right) \nonumber \\
    & & \times \exp\left[ - \int_0^1 dz\frac{dN_g^a}{dz}(Q_0^2)\right]\, ,
   \label{eq:pdeltaeps}
\end{eqnarray}
where $\langle N_g^a(Q_0^2)\rangle=\int_0^1 dzdN_g^a(Q_0^2)/dz$ is the average number of radiated gluons from the propagating parton ($a=q, g$).  Within the high-twist approach of parton energy loss, the induced gluon spectra per emission is given by
 \begin{eqnarray}
 \frac{dN_g^a}{dz}(Q_0^2) &=& \frac{\alpha_s(Q^2_0)}{2\pi}\int_0^{Q_0^2} \frac{d\ell_T^2}{\ell_T^2} \Delta\gamma_{a\rightarrow ag}(z,\ell_T^2), 
  \label{eq:dngdz}
  \end{eqnarray}
  where $\Delta\gamma_{a\rightarrow ag}(z,\ell_T^2)$ is the medium induced splitting function for parton $a$. For a quark jet  \cite{Schafer:2007xh}, for example,
  \begin{eqnarray}
 \Delta \gamma_{q\rightarrow qg}(z,\ell_T^2)&=&\frac{1}{\ell_T^2+\mu_D^2}\left[ C_{A}\frac{(1-z)(1+(1-z)^{2})}{z} \right. \nonumber \\
 &+& \left. C_{F}z(1+(1-z)^{2})\right] \nonumber \\
 &\times&  \int dy^-\hat q(y^-) 4 \sin^{2}(x_{L}p^{+}y^{-}/2),
 \label{delta_sp_qqg}
\end{eqnarray}
where $x_L=\ell_T^2/2p^+q^-z(1-z)$ is the fractional light-cone momentum of target partons that is required for the scattering to radiate a gluon, $y^-$ is the light-cone coordinate of the propagating parton, $\hat{q}(y^{-})$ is the quark transport 
parameter along the path and $\mu_D$ is a parameter representing gluon's average intrinsic transverse momentum inside a nucleon. In hot QGP, $\mu_D$ is replaced by the Debye screening mass. We freeze the running coupling constant at $\alpha_s(Q^2_0)$ below $Q_0^2$.  In the calculation of the averaged number of radiated gluons $\langle N_g^a(Q_0^2)\rangle$, we also impose kinematic constraints: $x_L\le 1$, $\ell_T^2/E^2\le z^2$ and $\ell_T^2/E^2\le (1-z)^2$.

Taking a Fourier transformation of the gluon spectra,
\begin{equation}
\widetilde N_g^a(\zeta, Q_0^2)=\int dz \frac{dN_g^a}{dz}(Q_0^2) e^{-iz\zeta},
\end{equation}
one can also cast the quenching weight in a compact form,
\begin{equation}
P_a(\epsilon, Q_0^2)=e^{-\langle N_g^a(Q_0^2)\rangle}\int \frac{d\zeta}{2\pi} e^{i\epsilon \zeta+\widetilde N_g^a(\zeta,Q_0^2)}.
\end{equation}

Numerical evaluation of the above quenching weight becomes difficult when the average number of emitted gluons  $\langle N_g^a\rangle$ is large. We have developed a Monte Carlo method to calculate the quenching weight $P_a(\epsilon,Q_0^2)$. This Monte Carlo method also provides more details about the total induced gluon spectra. In addition to $P_a(\epsilon,Q_0^2)$ which represents the probability of total fractional energy loss $\epsilon$ by the initial parton $a$ due to induced gluon radiation, we can also obtain $G^a(\epsilon)$ which represents the spectrum distribution of the radiated gluons with fractional energy $\epsilon$ from initial parton.  Note that $G^a(\epsilon,Q_0^2)$ is different from $dN_g^a(Q_0^2)/dz$. It is computed under the constraint that the total fractional energy loss via multiple gluon emissions by the initial parton can not be greater than one in each event. Because of momentum conservation, they should satisfy the momentum sum rule,
 \begin{eqnarray}
  \int_0^1 d\epsilon\,(1-\epsilon)P_a(\epsilon)+\int_0^1 d\epsilon \epsilon G^a(\epsilon)=1 \,.
   \label{eq:sumrule2}
\end{eqnarray}

With the quenching weights $P_a(\epsilon,Q_0^2)$ and the effective induced gluons' spectra $G^a(\epsilon,Q_0^2)$, one can obtain the modified fragmentation functions $\tilde D_{a}(z,Q_0^2)$ from a set of convolution equations:
 \begin{eqnarray}
  \tilde D_{h/g}(z,Q_0^2)&=&\int_0^1 d\epsilon P_g(\epsilon,Q_0^2)
  \frac{1}{1-\epsilon} D_{h/g}(\frac{z}{1-\epsilon},Q_0^2) \nonumber \\
  &+& \int_0^1 d\epsilon G^g(\epsilon,Q_0^2)\frac{1}{\epsilon}D_{h/g}(\frac{z}{\epsilon},Q_0^2)\,,\hspace{0.3in}   \label{eqquark}\\
  \tilde D_{h/q}(z,Q_0^2)&=&\int_0^1 d\epsilon P_q(\epsilon,Q_0^2)\frac{1}{1-\epsilon} D_{h/q}(\frac{z}{1-\epsilon},Q_0^2) \nonumber \\
  &+& \int_0^1 d\epsilon G^q(\epsilon,Q_0^2)\frac{1}{\epsilon} D_{h/g}(\frac{z}{\epsilon},Q_0^2).\hspace{0.3in} 
   \label{eqgluon}
\end{eqnarray} 
Note that the above mFF's  at the initial scale $Q_0^2$ include fragmentation of radiated gluons which ensures the total momentum conservation.  Using the momentum sum rules in Eq.~(\ref{eq:sumrule2}), one can verify that the above mFF's satisfy the momentum sum rules,
 \begin{equation}
\sum_h\int_0^1\mathrm{d}z z\tilde D_a^h(z,Q_0^2)=1, \,\,\, a=g,q,\bar q \, .
\end{equation}
Using these convoluted initial conditions as given by Eqs.~(\ref{eqquark}) and (\ref{eqgluon}), one can solve the mDGLAP equations in Eqs~(\ref{eq: modified DGLAP1}) and (\ref{eq: modified DGLAP2}) numerically and calculate the mFF's at any scale $Q^2$.

\section{The suppression factor in DIS}\label{sec:dis}

\begin{figure}
  \centering
  \includegraphics[width=0.4\textwidth]{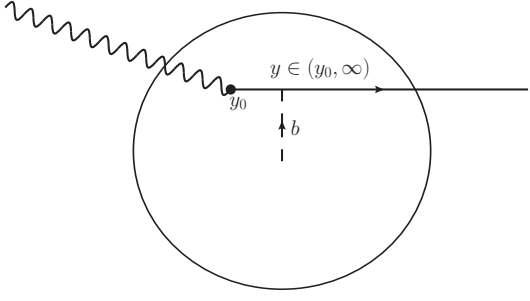}
  \caption{Illustration of the quark propagation path in DIS off a nucleus.}
  \label{fig:dis_plot}
\end{figure}

In semi-inclusive DIS off a nucleus, as illustrated in Fig.~\ref{fig:dis_plot}, a high energy virtual photon strikes out a quark from the nucleon at position $(y_0,b)$. The struck quark propagates through the rest of the nucleus along the path $y$ and loses energy due to multiple scatterings and induced gluon bremsstrahlung. The final quark and radiated gluons then fragment into hadrons. Jet quenching in the cold nuclear medium will be manifested in the nuclear modification of the final hadron spectra or jet fragmentation functions. In this section, we will calculate the modified fragmentation functions and the final hadron suppression factors in DIS using the convoluted initial condition and compare to the results with the vacuum initial condition and as well as the evolved initial condition as in our previous work \cite{Wang:2009qb,Deng:2010xv}.

The medium mFF's depend on the jet transport parameter in the medium through the modified splitting functions \cite{Wang:2009qb,Deng:2010xv} [see Eq.~(\ref{delta_sp_qqg}) for example] which in turn depend on the trajectory of the hard quark. One therefore needs to evaluate the path integration from the production point of the hard quark along the quark's trajectory in the modified splitting functions in the mDGLAP equations. One then has to average over the production point $(y_0,\vec b)$ weighted by the nucleon density inside the nucleus,
\begin{equation}
 \widetilde{D}(z,Q^2)=\frac{1}{A}\int d^2b dy_0 \widetilde{D}(z,y_0,b,Q^2) \rho_A(y_0,b),
\label{eq:average_dis}
\end{equation}
to obtain the averaged nuclear modified fragmentation functions.

We will employ the Woods-Saxon nuclear density distribution $\rho_{A}(y,b)$ which is normalized as $\int dy d^{2}b\rho_{A}(y,b)=A$. We also assume the jet transport parameter $\hat{q}$ along the quark jet trajectory is proportional to the local nuclear density, 
\begin{equation}
\hat q(y,b)=\hat q_{0} \frac{\rho_{A}(y,b)}{\rho_{A}(0,0)},
\end{equation}
where $\hat q_{0}$ is defined to be the value of $\hat q$ at the center of the nucleus.

In HERMES experiment \cite{HERMES}, ratios between the hadron multiplicities from a nucleus target and that from deuteron are measured, which
can be expressed in terms of the modified fragmentation functions,
\begin{eqnarray}
R_A^h(z,\nu)&=&\left(\frac{N^h(z,\nu)}{N^e(\nu)}|_A\right) / \left(\frac{N^h(z,\nu)}{N^e(\nu)}|_D\right) \nonumber\\
& & \hspace{-0.7in} =\left(\frac{\Sigma e_q^2q(x)\widetilde D_q^h(z)}{\Sigma e_q^2q(x)}|_A\right) / \left(\frac{\Sigma e_q^2q(x) D_q^h(z)}{\Sigma e_q^2q(x)}|_D\right),
\end{eqnarray}
where $\nu=E$ is the energy of virtual photon that is transferred to the struck quark , $z=p_h/E$ is the energy fraction carried by the final hadrons, $x=Q^2/2p^+q^-=Q^2/2M_N\nu$ the Bjorken variable for fractional light-cone momentum carried by the initial quark. The summation is over all quark and anti-quark flavors and $q(x)$'s are the quark distributions inside the nucleus. We use the CTEQ6 parameterization \cite{Pumplin:2002vw} of parton distributions and EKS parameterization \cite{EKS} of the nuclear modification of the parton distributions.

In Fig.~\ref{fig-rhz} we compare our calculated results on the suppression factor $R_A^h$ in DIS (lines) off three different targets with the convoluted initial condition  with the HERMES data \cite{HERMES} as a function of the hadrons' final fractional momentum $z$. We have choosen $Q_0=1$ GeV and set $\mu_D=0.2$ GeV  which is related to a gluon's average intrinsic transverse momentum inside a nucleon. We use the HKN parametrization \cite{HKN}  for the vacuum fragmentation functions at $Q_0^2$. We also use the corresponding averaged values of $Q^2$ and $\nu$ for each bin of $z$ according to that in the HERMES experiment. The calculated results agree with the HERMES data quite well for pions and kaons for small and intermediate values of $z$. 
At large values of $z$ the agreement is not so good, possibly due to other effects such as hadronic interaction \cite{hadronic,Arleo:2003jz} that are not considered in our study.  The theoretical results also over-estimate the suppression for protons and under-estimate the suppression for anti-protons. This might be related to the non-perturbative baryon transport in hadronic processes \cite{Kharzeev:1996sq} and hadronic interaction since baryons' formation can be shorter than that for pions and kaons. We have also neglected quark- anti-quark annihilation contribution to the mDGLAP evolution equations. These processes will affect the medium modification of anti-quarks and will likely improve the modification factor for anti-proton spectra.

\begin{widetext}
\begin{center}
\begin{figure}[htb]
  \centering
     \includegraphics[width=6.0in]{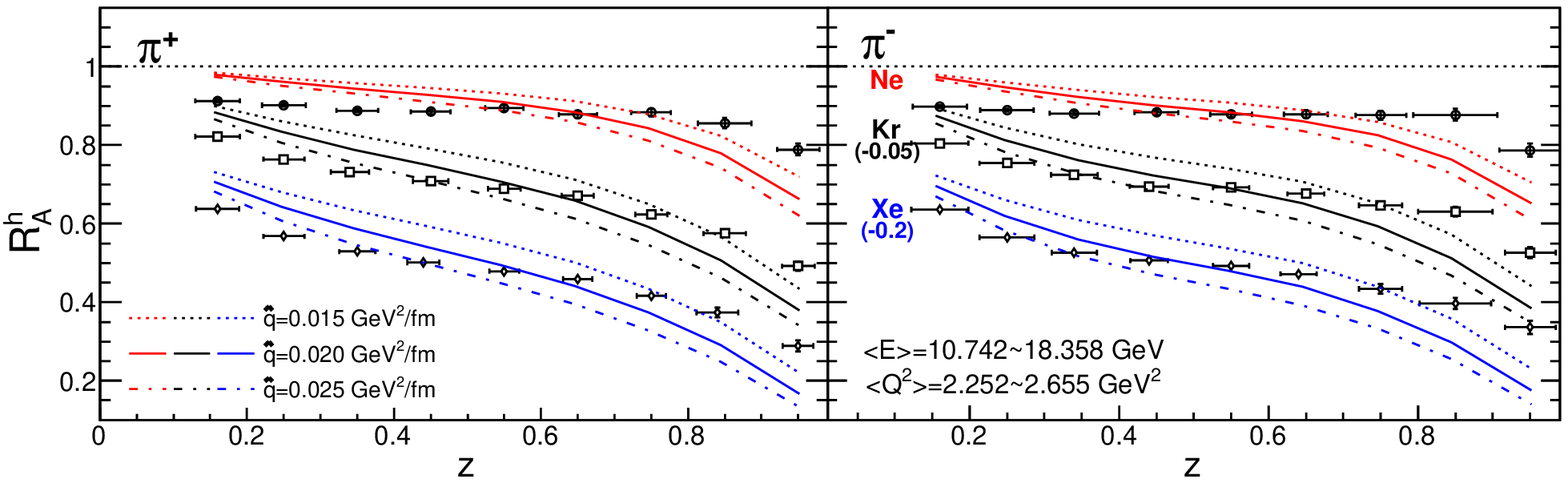}
     \includegraphics[width=6.0in]{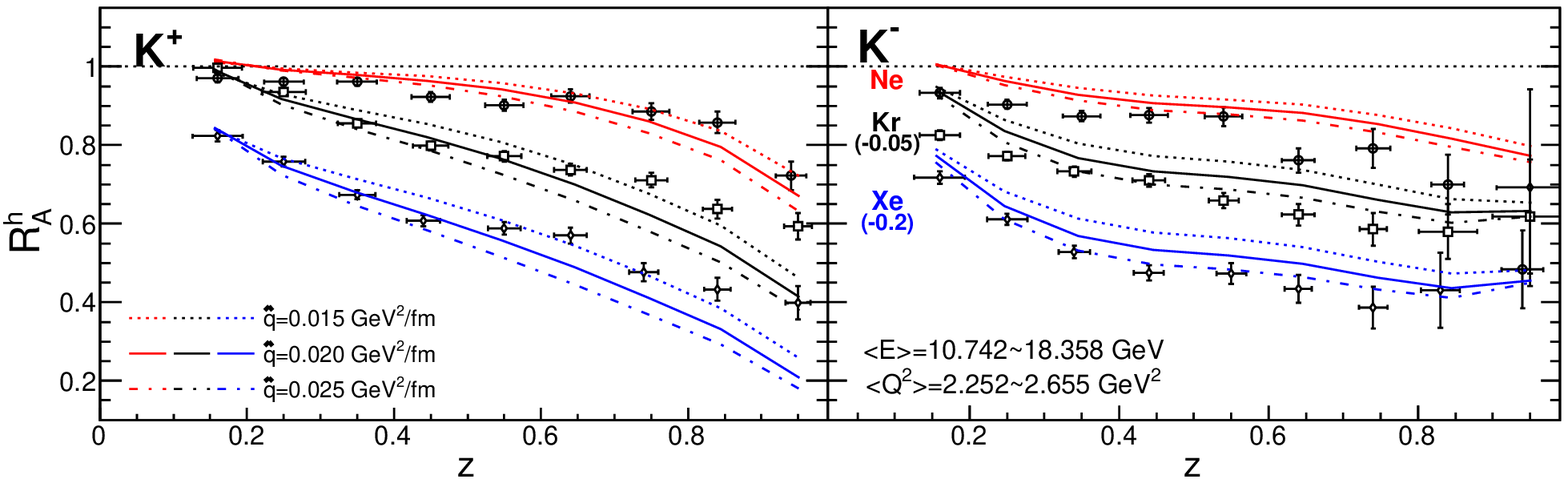}
     \includegraphics[width=6.0in]{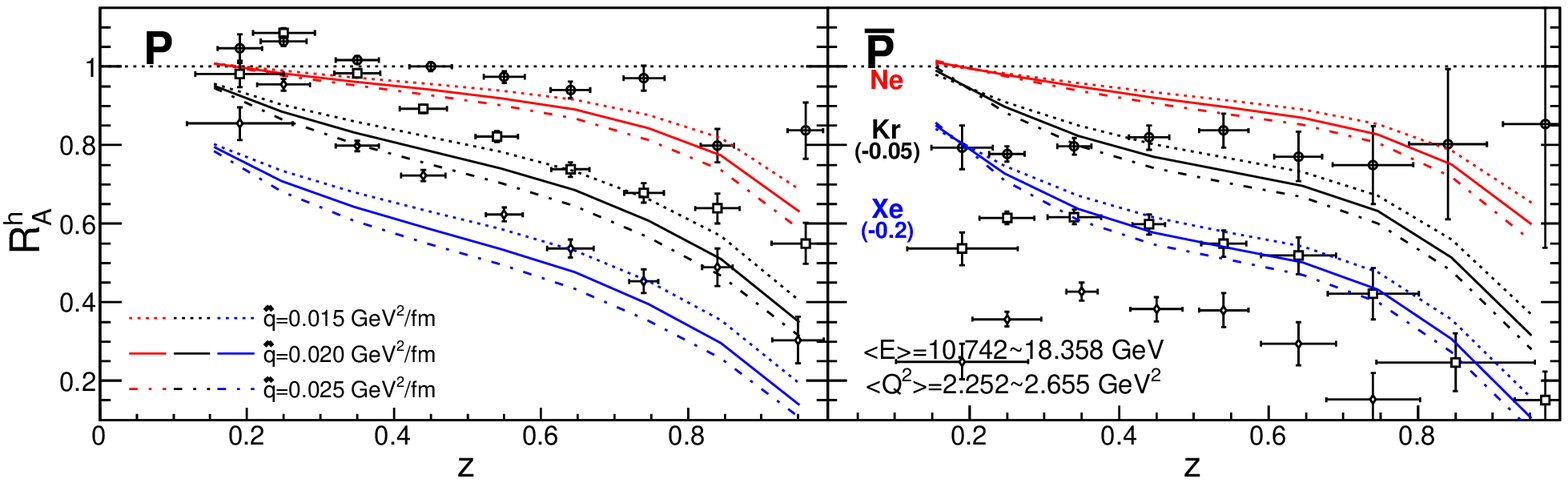}
   \caption{(color online) The $z$ dependence of calculated $R^h_A$  for pions (top), kaons (middle), protons and anti-protons (bottom panel) with the convoluted initial condition for different values of $\hat{q}_0$ compared with HERMES data\cite{HERMES} for Ne, Kr and Xe targets. For clarity, values of $R_A^h$ for Kr and Xe targets are displaced by -0.05 and -0.2, respectively.}
  \label{fig-rhz}
\end{figure}
\begin{figure}[htb]
  \centering
     \includegraphics[width=6.0in]{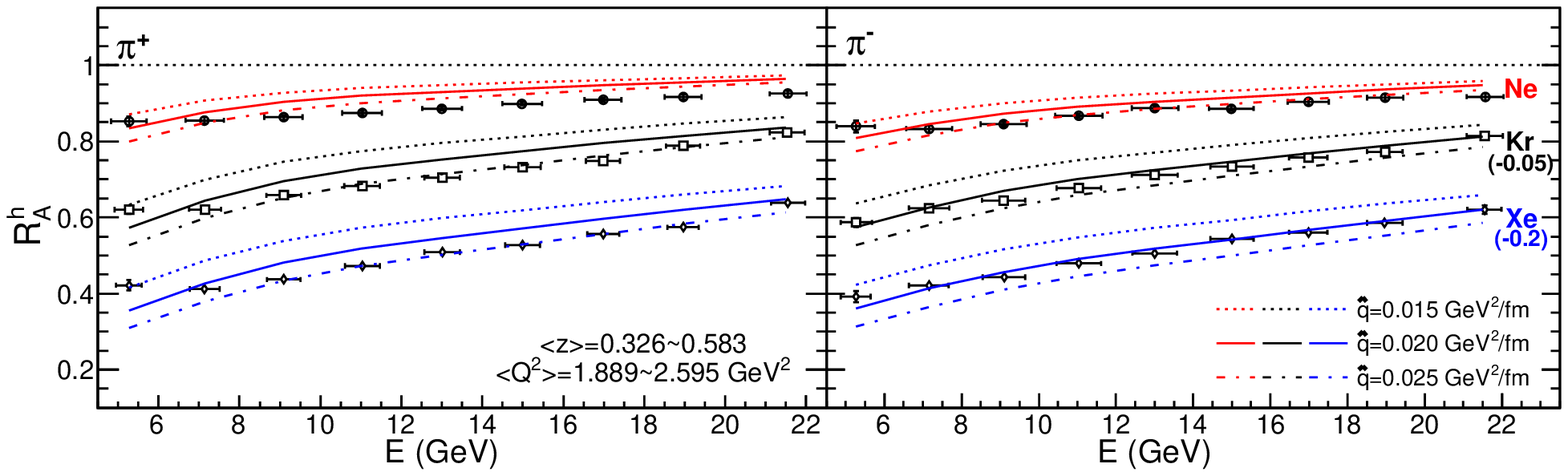}
     \includegraphics[width=6.0in]{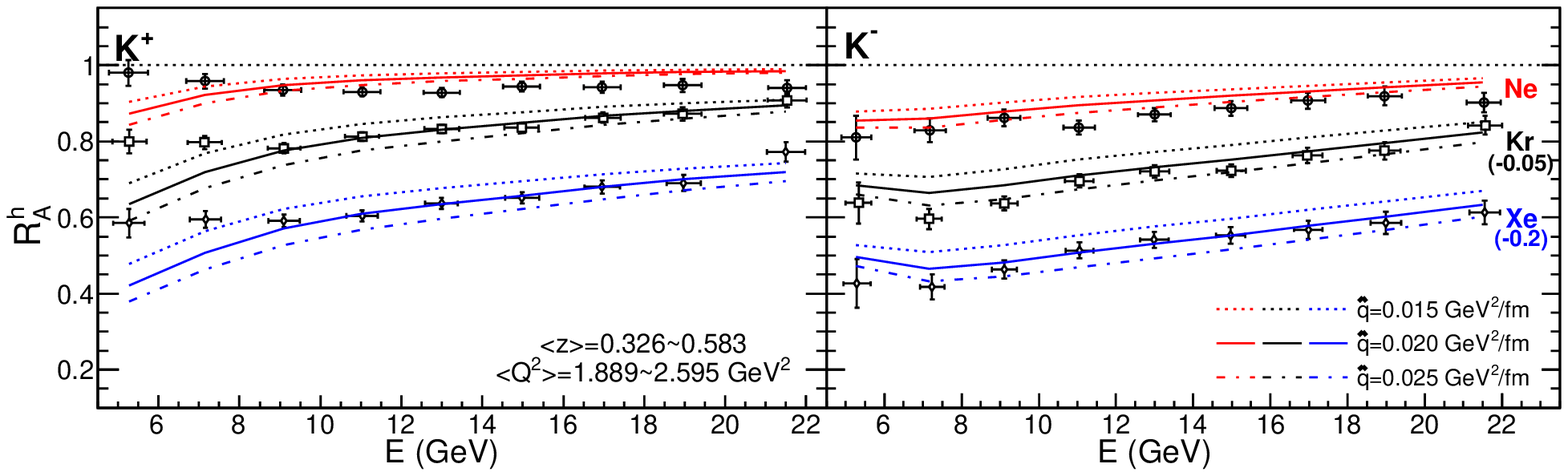}
     \includegraphics[width=6.0in]{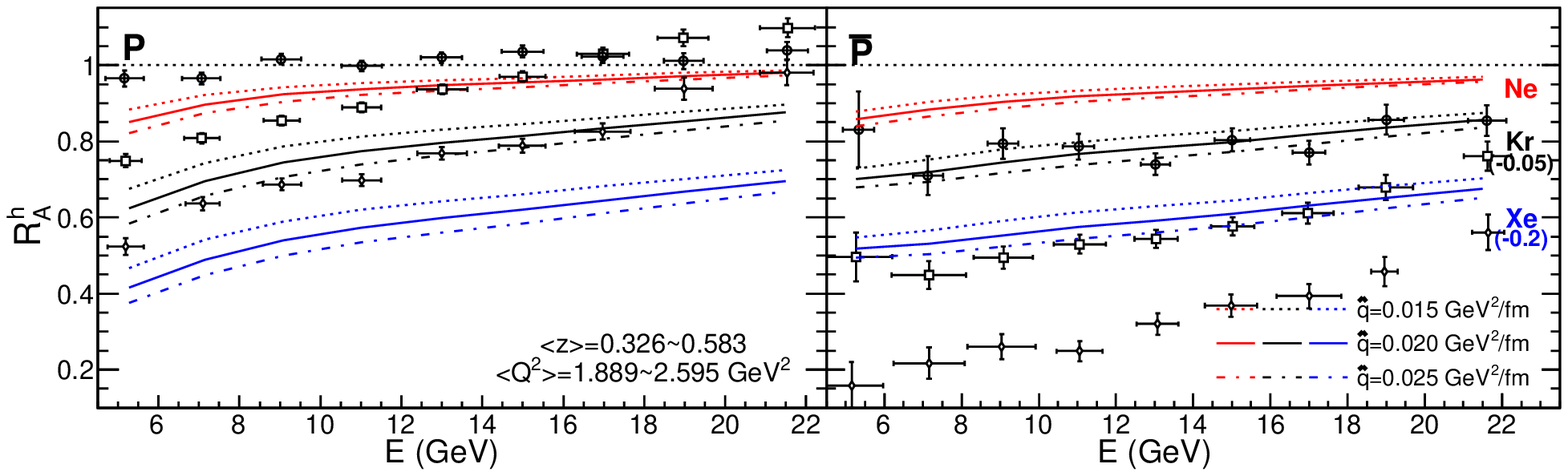}
   \caption{(color online)  The same as Fig.~\ref{fig-rhz} except for the suppression factor as a function of initial quark energy $E$.}
  \label{fig-rhe}
\end{figure}
%
\begin{figure}[htb]
  \centering
     \includegraphics[width=6.0in]{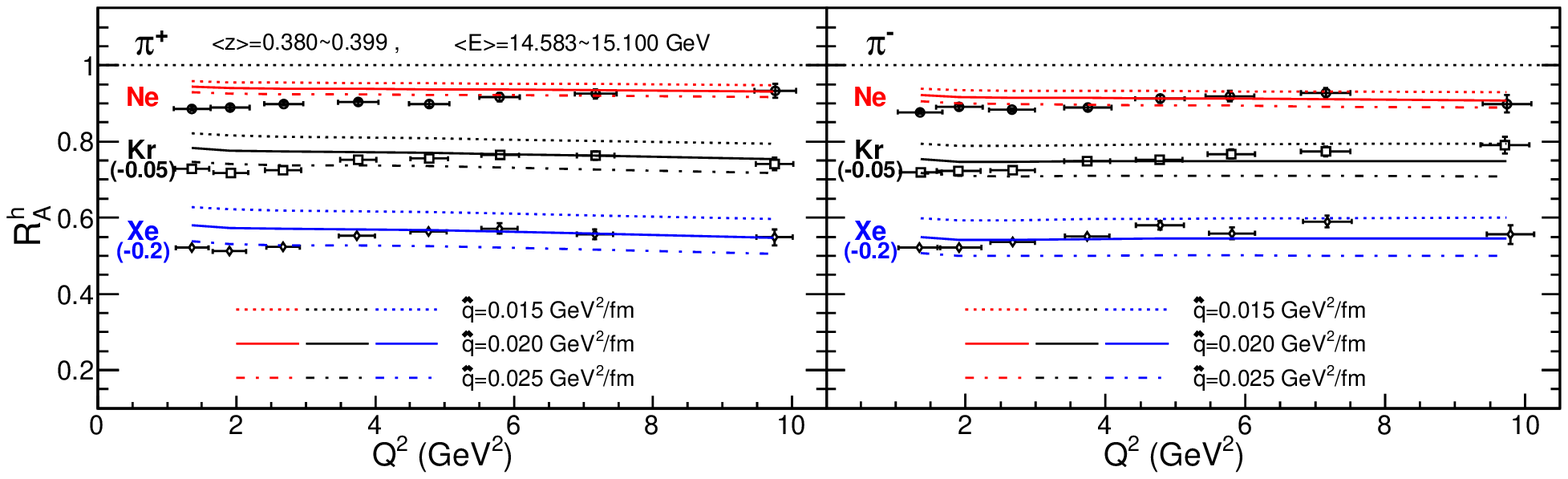}
     \includegraphics[width=6.0in]{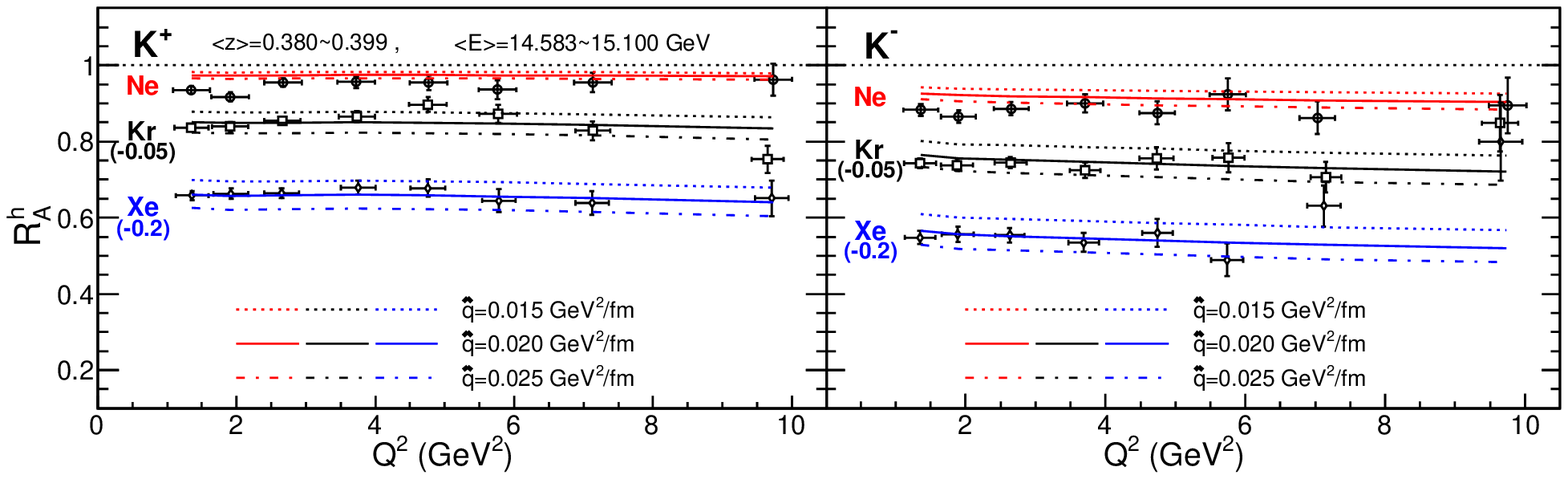}
   \caption{ (color online) The same as Fig.~\ref{fig-rhz} except for the suppression factor as a function of initial quark virtuality $Q^2$.}
  \label{fig-rhq}
\end{figure}
\end{center}
\end{widetext}

We can also calculate $R_A^h$ as a function of initial quark energy $E$ for a given range of $z$ and $Q^2$. Again, the range of $z$ and $Q^2$ varies with the value of $E$. The results are compared with the HERMES experimental data in Fig.~\ref{fig-rhe}. The agreement is again very good for pions and kaons but not so good for protons and anti-protons.  Shown in Fig.~\ref{fig-rhq} are the suppression factors as a function of $Q^2$ which are quite weak as also indicated by the HERMES data.

\begin{figure}[htb]
  \centering
     \includegraphics[width=3.0in]{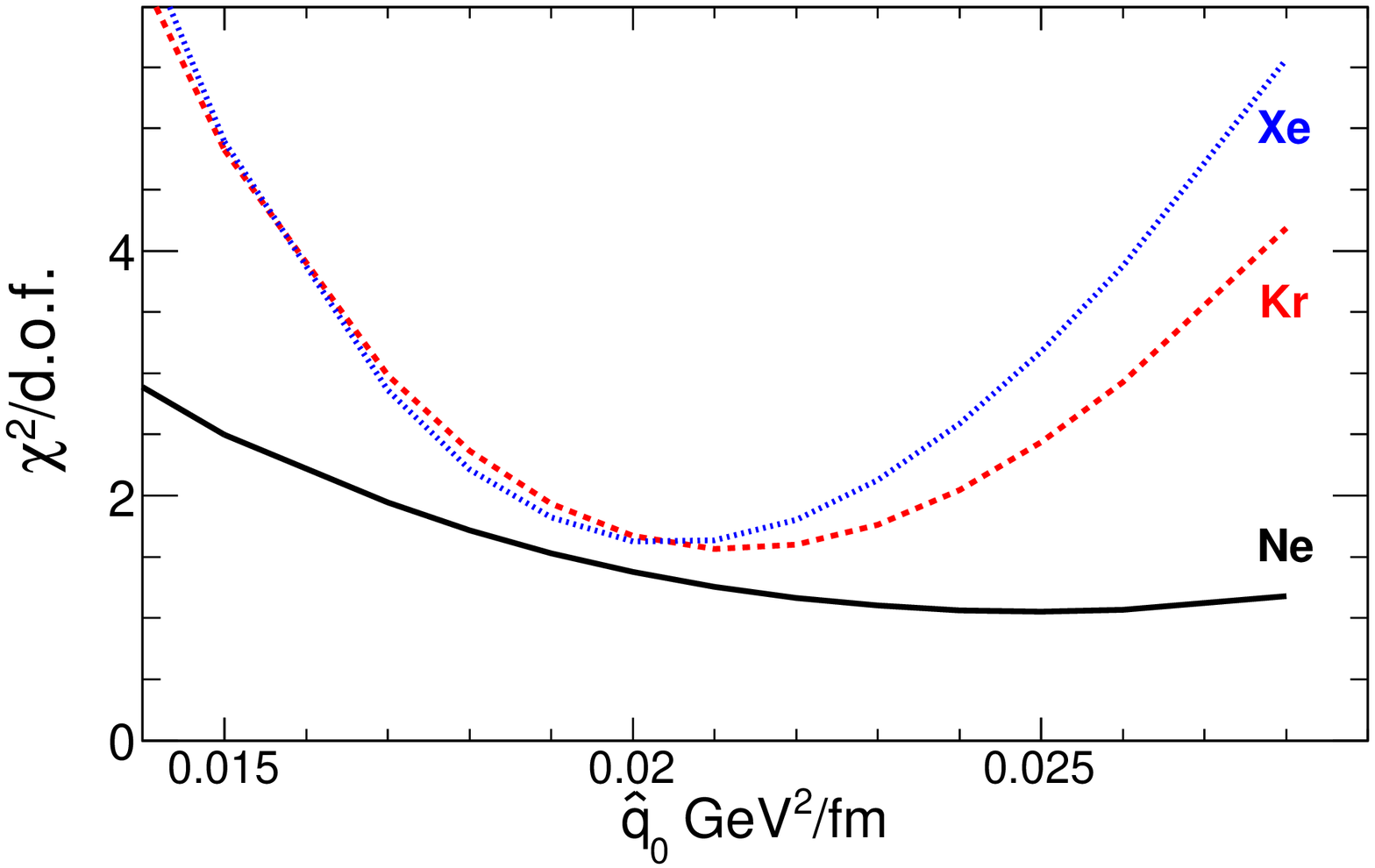}
     \includegraphics[width=3.0in]{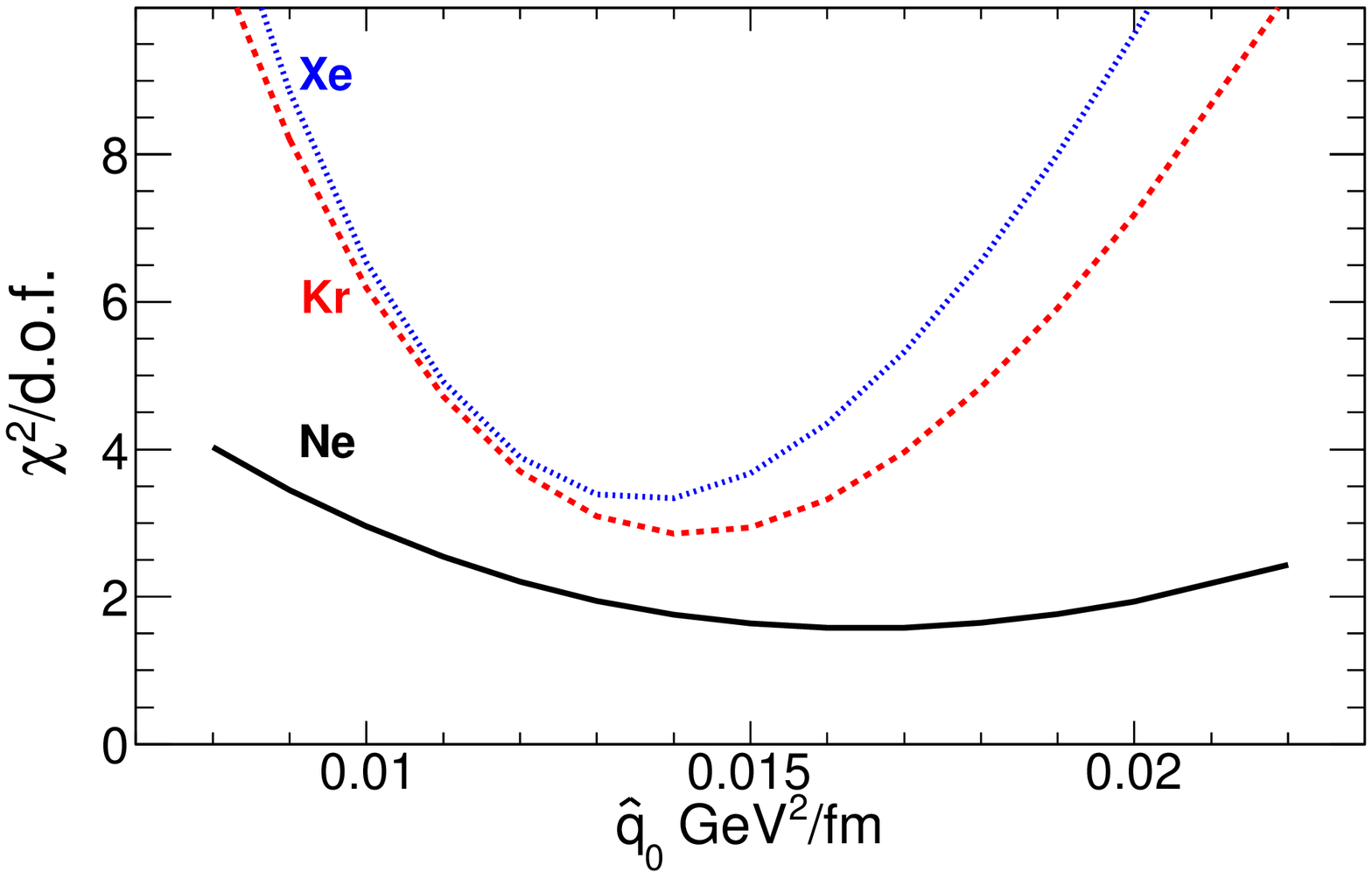}
      \includegraphics[width=3.0in]{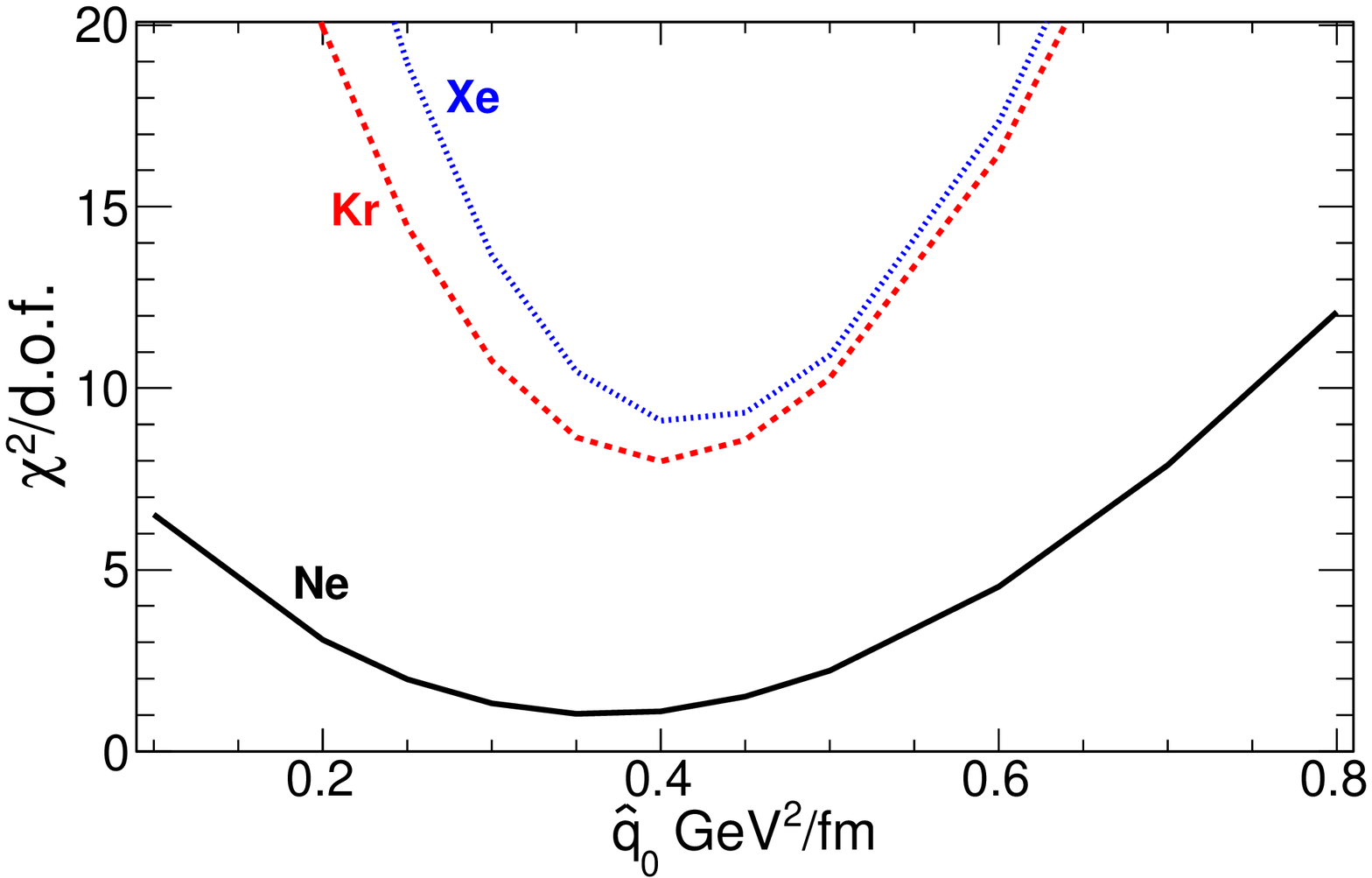}
   \caption{(color online) $\chi^2$/d.o.f. as a function of $\hat q_0$ from fits to the HERMES data \cite{HERMES}  with the calculated results on the $E$-dependence of the suppression factors from mDGLAP evolution equations using convoluted (top), evolved (middle) and vacuum initial conditions (bottom panel).}
  \label{chi2}
\end{figure}

\begin{figure}[htb]
  \centering
     \includegraphics[width=3.0in]{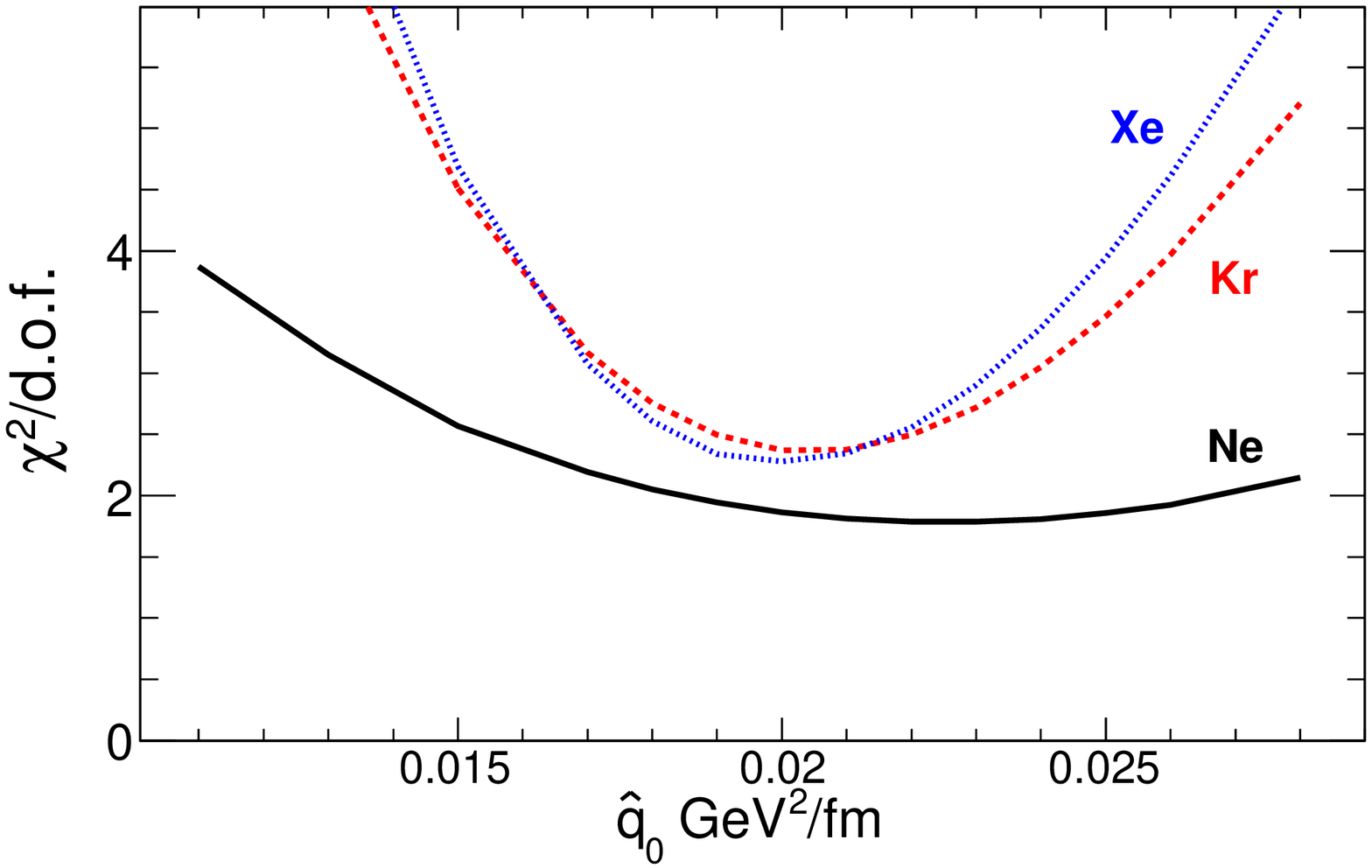}
     \includegraphics[width=3.0in]{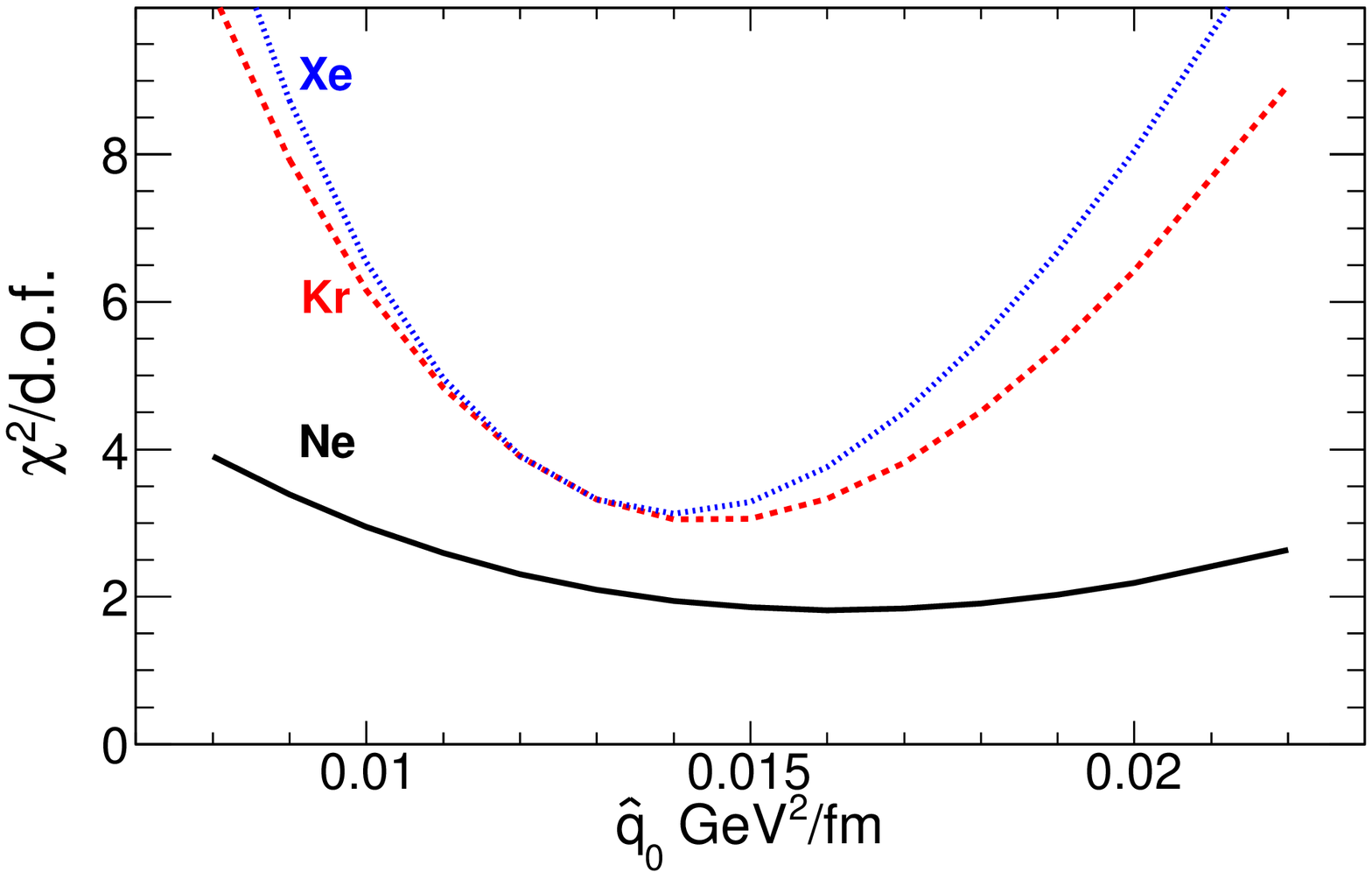}
      \includegraphics[width=3.0in]{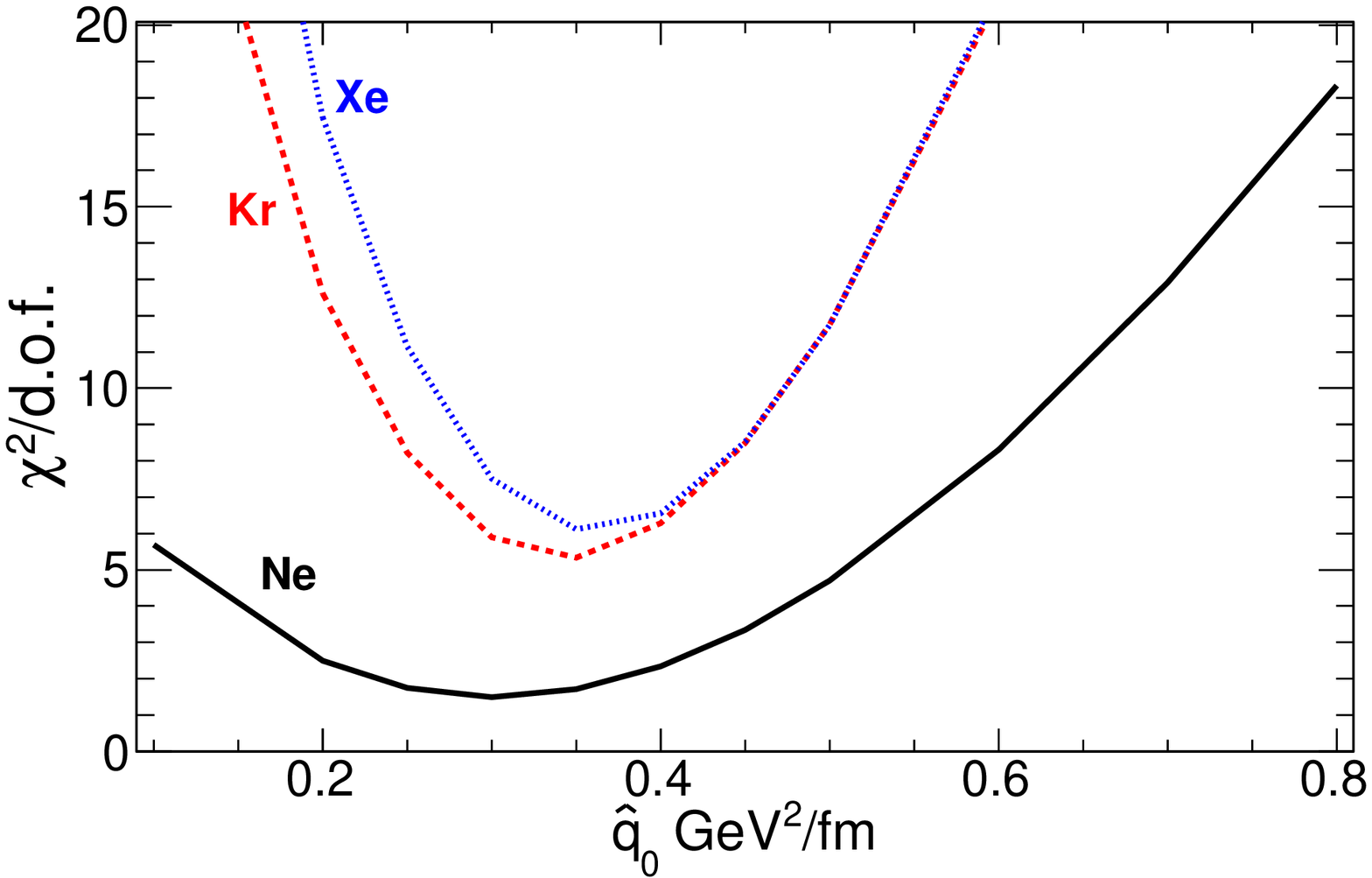}
   \caption{(color online) The combined $\chi^2$/d.o.f. as a function of $\hat q_0$ from fits to the HERMES data \cite{HERMES} with the calculated results on the $z$, $Q^2$ and $E$-dependence of suppression factors from mDGLAP evolution equations using convoluted (top), evolved (middle) and vacuum initial conditions (bottom panel).}
  \label{chi2-comb}
\end{figure}

To illustrate the quality of the fits to experimental data as compared to early models for initial conditions for the mDGLAP evolution and extract the best values of jet transport parameter $\hat q_0$, we plot in Fig.~\ref{chi2} $\chi^2$/d.o.f. of the fits to the HERMES data on the quark energy $E$-dependence of the suppression factor $R_A^h(z,E,Q^2)$  as a function of $\hat q_0$ for calculations with the convoluted (top), evolved (middle) and vacuum (bottom panel) initial conditions. The HERMES data used in the fits are only for the suppression factors of pions and kaons since the proton and anti-proton suppressions are complicated by other mechanism beyond the high-twist framework. We can see that the convoluted initial condition proposed in this paper gives the smallest values of  $\chi^2$/d.o.f. at the minima as compared to the evolved and vacuum initial conditions. The vacuum initial condition gives the largest values of $\chi^2$/d.o.f. at the minima.  Similar $\chi^2$/d.o.f. analysis of the fits to the HERMES data on the $Q^2$-dependence of the suppression factors $R_A^h(z,E,Q^2)$ prefers the convoluted initial conditions over the evolved and vacuum ones while the vacuum one again is the least preferred. However, the vacuum initial condition is found to fit the data on the $z$-dependence of the suppression factors slightly better than the evolved and convoluted ones.  Shown in Fig.~\ref{chi2-comb} are $\chi^2$/d.o.f. from combined fits to $z$, $E$ and $Q^2$-dependence of the suppression factors from the HERMES experiment. The convoluted initial condition proposed in this paper has the best overall fit to the data among the three initial conditions we have studied and gives a fitted value $\hat q_0=0.020\pm 0.005$ for the quark transport parameter at the center of large nuclei.

Among the three initial conditions, the vacuum one has the worst fit to the experimental data.  With the vacuum initial condition, the medium modification of the fragmentation functions solely comes from the mDGLAP evolution and therefore has much stronger $Q^2$ dependence as shown in Fig.~\ref{fig-rhq-vac}. It also requires a large value of $\hat q_0$ to give a large suppression as expected. One can improve the $Q^2$ dependence by introducing a propagation length dependence in the value of $Q_0^2$ \cite{Majumder:2011uk} which could be quite large than the value of 1 GeV$^2$ we used here.

Note that our analyses here are based on the high-twist approach in which only contributions in leading order (LO) pQCD up to twist-four are considered currently. The jet transport parameter $\hat q$ is considered a constant in this case. In principle, $\hat q$ should depend on jet energy $E$ and virtuality $Q^2$ \cite{CasalderreySolana:2007sw} due to multiple  gluon emission processes associated with the target partons in medium. Such energy and virtuality dependence will arise naturally when next-to-leading order (NLO) corrections at twist-four are considered \cite{Mueller:2012bn,Liou:2013qya,Kang:2013raa}. These NLO corrections should be considered for further improvements in jet quenching studies.

\begin{widetext}
\begin{center}
\begin{figure}[htb]
  \centering
     \includegraphics[width=6.0in]{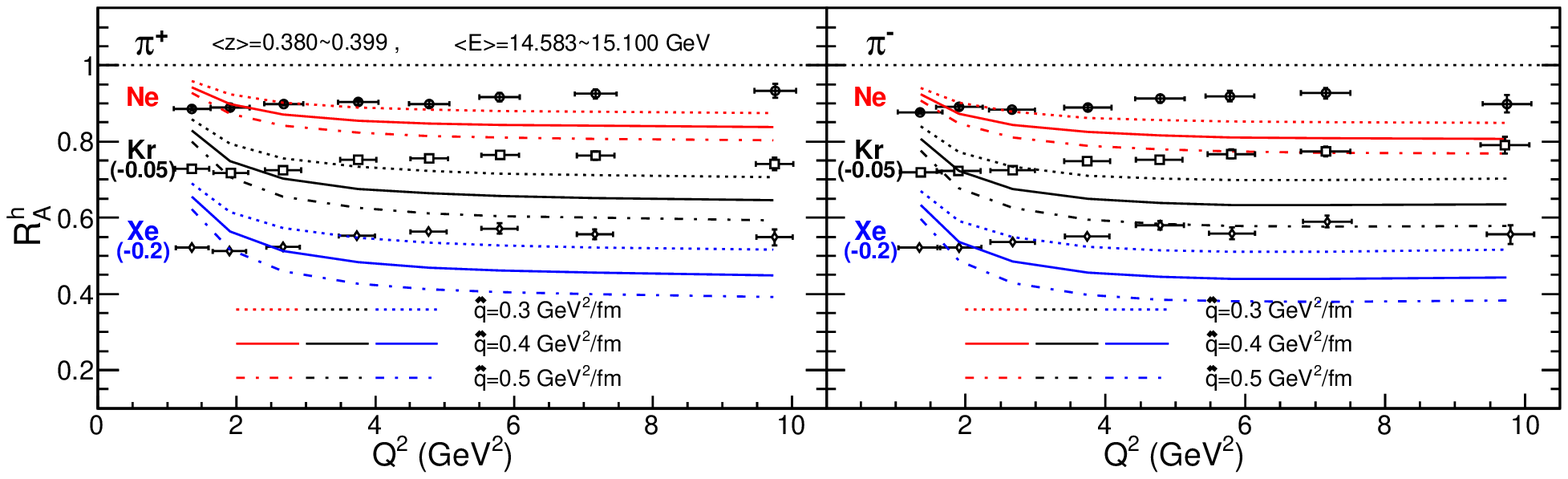}
        \includegraphics[width=6.0in]{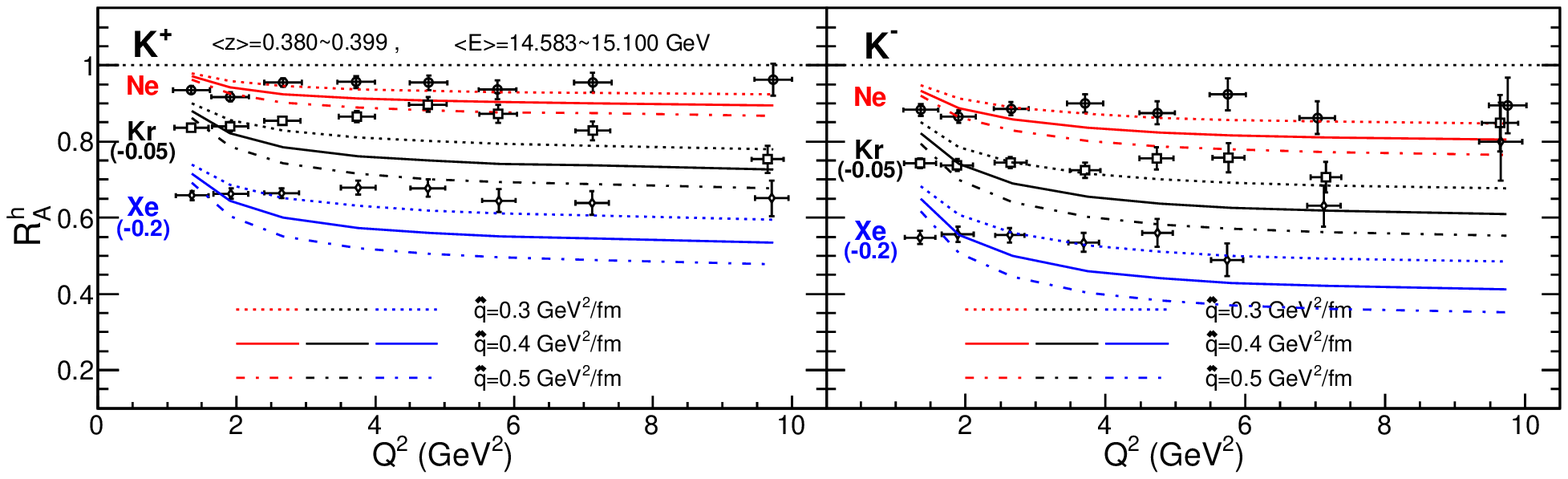}
   \caption{ (color online) The $Q^2$ dependence of calculated $R^h_A$  for pions (top) and kaons (bottom panel) with the vacuum initial condition for different values of $\hat{q}_0$ compared with HERMES data\cite{HERMES} for Ne, Kr and Xe targets.}
  \label{fig-rhq-vac}
\end{figure}
\end{center}
\end{widetext}

\section{Summary} \label{sec:summ}

In summary, we have studied the medium modification of fragmentation functions through a set of modified DGLAP evolution equations within the high-twist approach with different initial conditions. We proposed a convoluted initial condition which is a Poisson convolution of multiple gluon radiations each has a spectrum from a single gluon emission. By fitting to the experimental data on hadron suppression factors in DIS off nuclei, we find that the convoluted initial condition gives the best $\chi^2$/d.o.f. fit as compared to the evolved and vacuum initial conditions that were used in previous studies.  Such convoluted initial conditions can also be used to study jet quenching in high-energy heavy-ion collisions. The value of jet transport parameter in cold nuclear matter $\hat q_0$ extracted in this study will also provide improved model for jet transport parameter in a hadron resonance gas at finite temperature as part of the jet quenching mechanism throughout the evolution history of the dense matter in high-energy heavy-ion collisions.


{\bf Acknowledgments:} 
This work was supported by the National Natural Science Foundation of China
under the grant No. 11221504 and 11035003, the Major State Basic Research Development Program in China (No. 2014CB845404) and by the Director, Office of Energy Research, Office of High Energy and Nuclear Physics, Division of Nuclear Physics, of the U.S. Department of  Energy under Contract No. DE-AC02-05CH11231 and within the framework of the JET Collaboration. W.-T. Deng was supported in part by Grant-in Aid for Scientific Research (No. 22340064) from the Ministry of Education, Culture, Sports, Science and Technology (MEXT) of Japan. N.-B. Chang was support in part by CCNU-QLPL Innovation Fund (QLPL2011P01).


\end{document}